# A microrod-resonator Brillouin laser with 240 Hz absolute linewidth

William Loh*, Joe Becker, Daniel C. Cole, Aurelien Coillet, Fred N. Baynes, Scott B. Papp and Scott A. Diddams

*National Institute of Standards and Technology, 325 Broadway, Boulder, Colorado 80305, USA*
*William.Loh@nist.gov

**Abstract:** We demonstrate an ultralow-noise microrod-resonator based laser that oscillates on the gain supplied by the stimulated Brillouin scattering optical nonlinearity. Microresonator Brillouin lasers are known to offer an outstanding frequency noise floor, which is limited by fundamental thermal fluctuations. Here, we show experimental evidence that thermal effects also dominate the close-to-carrier frequency fluctuations. The 6-mm diameter microrod resonator used in our experiments has a large optical mode area of ~100 μm$^2$, and hence its 10 ms thermal time constant filters the close-to-carrier optical frequency noise. The result is an absolute laser linewidth of 240 Hz with a corresponding white-frequency noise floor of 0.1 Hz$^2$/Hz. We explain the steady-state performance of this laser by measurements of its operation state and of its mode detuning and lineshape. Our results highlight a mechanism for noise that is common to many microresonator devices due to the inherent coupling between intracavity power and mode frequency. We demonstrate the ability to reduce this noise through a feedback loop that stabilizes the intracavity power.

## 1. Introduction

Whispering gallery mode microresonators have emerged as a system that can couple light and matter at extremely low levels of loss [1–6]. In recent years, this technology has evolved to the point where intrinsic quality factors (Q) of $10^9$ can be achieved in a size scale of a few millimeters or less. The high Q of these resonators significantly enhances the optical power that circulates within the cavity and thus provides a path for realizing low-power nonlinear optics in a compact form. With the recent advances in the microresonator technology, optical frequency combs based on the four-wave mixing process [7–11] and low-noise lasers based on the stimulated Brillouin [12–14] or stimulated Raman [15] processes have become a reality.

Traditionally, the stimulated Brillouin scattering (SBS) process [16] has been utilized as a means to generate low-noise lasing in bulk-fiber ring cavities [17–20]. These Brillouin lasers take advantage of the low optical losses in optical fiber and the narrow bandwidth of the SBS gain to achieve single-mode lasing with exceptionally high levels of spectral purity [21, 22]. Recently, the development of ultra-high Q microcavities has enabled a new regime of low intracavity loss capable of supporting the generation of nonlinear optics at optical pump powers below 1 mW. These microcavity SBS lasers operate with ultra-narrow linewidths based on their ability to simultaneously achieve low levels of thermally-excited noise while still maintaining large signal powers, thus yielding the largest ratios of signal to noise in any laser. With the use of centimeter-scale microresonators, the Brillouin laser currently achieves linewidths as low as a few kilohertz [14]; however further research may eventually enable linewidths that reach the fundamental Schawlow-Townes limit of < 1 Hz. Such lasers are expected to find application in miniaturizing systems for low-noise optical frequency division [23], ultra-stable optical atomic clocks [24], precision spectroscopy [25], and high-resolution interferometry [26].

The narrowest linewidth so far in Brillouin microcavity lasers of < 90 Hz was achieved in a silica microdisk SBS laser by frequency stabilizing the SBS laser output to a high-Q fused silica microrod reference cavity [14]. The large mode volume of the microrod reference cavity reduced thermorefractive noise at low offset frequencies, while for higher frequencies, the intrinsic noise of the SBS process resulted in a low value of 1 $Hz^2/Hz$ for the broadband white-noise floor. When operated in combination, the system exhibited low noise across the entire range of the lasing spectrum, at a level 2-3 orders of magnitude below that of commercial Er fiber lasers. This approach mimics the traditional route to achieving narrow linewidth lasers in which the active laser is separate from its optical reference.

The goal of this work is to understand and control the underlying noise limitations in order achieve ultra-narrow linewidth SBS lasing without the aid of an external optical reference microcavity. Toward this goal, in the following sections we present both theory and experiment to analyze the operation state of a Brillouin microcavity laser. Our experimental results show that the SBS laser's amplitude fluctuations become coupled to its frequency fluctuations through the thermal response of the microcavity. A different form of this coupling governing the interaction between the frequency of the pump laser and the microcavity temperature was also the subject of a previous study in Ref. [27]. These measurements suggest that the use of a large optical mode volume microresonator, such as our silica microrod, would increase the cavity's thermal time constant and offer reduction in the SBS laser's noise. By generating SBS lasing in a microrod resonator, which has a mode area of approximately 100 $\mu m^2$, we demonstrate two orders of magnitude reduction in close-to-carrier frequency noise compared to a silica microdisk resonator [14], thus reaching linewidths of < 240 Hz. As a point of comparison, the lasing linewidth of our centimeter-scale SBS laser is at least 10–100 times smaller than what is typically achieved in the lowest-noise fiber, diode and solid state lasers [28]. In addition, we utilize the coupling between amplitude and frequency as a means to stabilize the SBS laser frequency based on a measurement of intensity noise. This servo further reduces our laser noise by 10 dB and correspondingly decreases the laser linewidth to 225 Hz.

## 2. Results

### A. SBS Laser Setup

The configuration of our SBS microcavity laser is illustrated in Fig. 1a and consists of a commercial integrated planar external-cavity diode laser (pump) that supplies optical power to a microrod resonator via the evanescent field coupling of a tapered fiber. The resulting generated SBS wave is collected through a circulator (CIR) in the counter-propagating direction. A semiconductor optical amplifier (SOA) is used for the dual purpose of amplifying the pump power to 70.8 mW and for allowing independent control of the pump amplitude with 100 kHz bandwidth. The electro-optic phase modulator (PM) combined with the photodetector (PD) provides the ability to Pound-Drever-Hall lock our pump laser to the cavity resonance while simultaneously granting access to the pump laser phase. Although the total pump power supplied is 70.8 mW, the optical power delivered to the tapered fiber input is 11.7 mW, due to both the losses in the system and to the output couplers required for signal monitoring. With 11.7 mW input, the generated SBS signal reaches a level of 3.7 mW in the steady state. The corresponding spectrum of the SBS signal is shown in Fig. 1b operated at a center wavelength of 1550 nm. A small amount of backscattered pump power is also apparent in Fig. 1b separated from the SBS signal by 11.2 GHz, and can be removed through the use of additional filtering if necessary. We found this level of backscattered pump to be above the residual reflections of the system obtained by removing the microresonator from the setup.

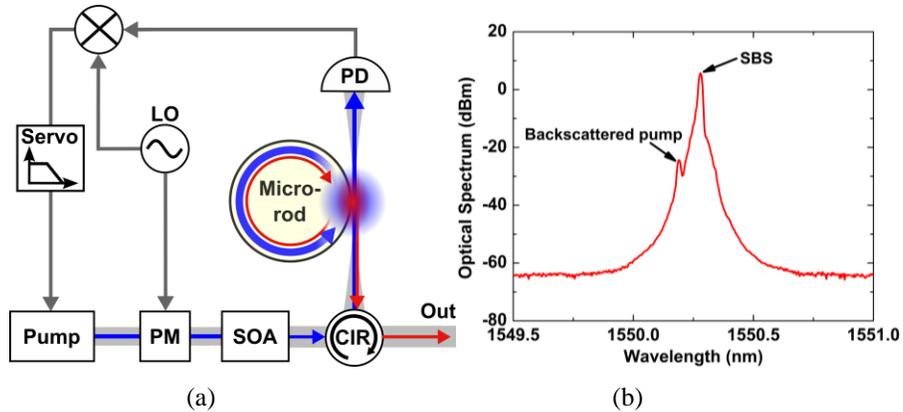

(a)        (b)

**Figure 1.** SBS microrod laser. (a) Configuration of the SBS laser that employs a commercial integrated planar external-cavity diode laser to pump a high-Q microrod cavity. (b) The output optical spectrum of the SBS laser exhibiting 3.7 mW power at 1550 nm wavelength.

## B. SBS Gain and Microcavity Characterization

Knowledge of the SBS laser operation state equips us with the ability to understand and diagnose the laser's performance. We characterize the operation of the SBS laser through measurements of the linewidth and detuning for both the pump and SBS modes. Our measurements are performed in conjunction with the SBS signal generation process so that we effectively measure the laser operation under conditions when the cavity is loaded. A diagram of our measurement setup is shown in Fig. 2a. The pump laser is stabilized onto the cavity resonance peak for the generation of SBS, while a tunable probe laser is sent in the backwards direction to measure the cavity lineshape. The scan over the pump mode is provided in Fig. 2b and depicts both the cavity resonance and also a residual undulation due to the beating of the probe laser with the stray reflected pump light [29]. As can be observed in Fig. 2b, the beat frequency approaches zero near the peak of the resonance, verifying that our pump laser is locked to the microcavity with nearly zero detuning. A Lorentzian fit to the measured mode profile yields a corresponding linewidth of 2.0 MHz. Note that we calibrate the frequency axis of Fig. 2b by phase modulating the probe laser at 10 MHz and subsequently matching the resulting separation of the sidebands when swept over the mode as measured by an oscilloscope to the known frequency spacing of the sidebands.

Figure 2c shows the profile of the microresonator mode used for the generation of SBS along with the position of the SBS signal relative to the mode resonance. Unlike the case of the pump mode scan in Fig. 2b, the strength of the counter-propagating SBS signal is much larger in the backwards direction, and the corresponding beating with the probe signal would dominate the scan over the SBS mode. For this reason, we instead phase modulate our pump laser to create sidebands at 11.2 GHz and sweep the modulation sidebands over the SBS mode in the forward-propagating direction. Since the dispersion in the microrod cavity generally shifts the resonances by only a few kilohertz per cavity mode, the scan of the modulation sidebands over the SBS mode inevitably averages over the profile of the mode on the opposite side of the pump. However, since all the modes are within the same mode family and are closely spaced in frequency, we do not expect significant distortion in the measurement of the SBS mode. In addition to measuring the mode profile, we also combine our forward- and reverse-propagating paths on a photodetector and use the resulting beating to measure the separation of the SBS signal from the pump. Putting these measurements together, we find the SBS signal to be blue-detuned by 810 kHz compared to the cavity resonance. Note that the asymmetry observed in the mode profile of Fig. 2b is due to the presence of higher-order microrod spatial modes near the SBS resonance.

The measurements of linewidth and detuning for the pump and SBS modes yield considerable insight regarding the operating state of the SBS laser. We now show that these measurements provide us with information on the SBS gain bandwidth and detuning as well as the lifetime and relative phase of the photons in the microcavity. This knowledge provides the foundation for understanding the performance of the SBS laser in terms of its output power, noise, and linewidth. From the necessity to satisfy energy and momentum conservation, the steady-state operation of the laser enforces that the pump, SBS, and phonon mode detuning are related by [22]

$$\sigma_B = \frac{\sigma_F - \left[\left(\Omega_b^2 - \Omega^2\right)/2\Omega\right]}{1+\Gamma_b\tau_B}$$
$$\sigma_\rho = \frac{\sigma_F \Gamma_b \tau_B + \left[\left(\Omega_b^2 - \Omega^2\right)/2\Omega\right]}{1+\Gamma_b\tau_B} \quad (1)$$

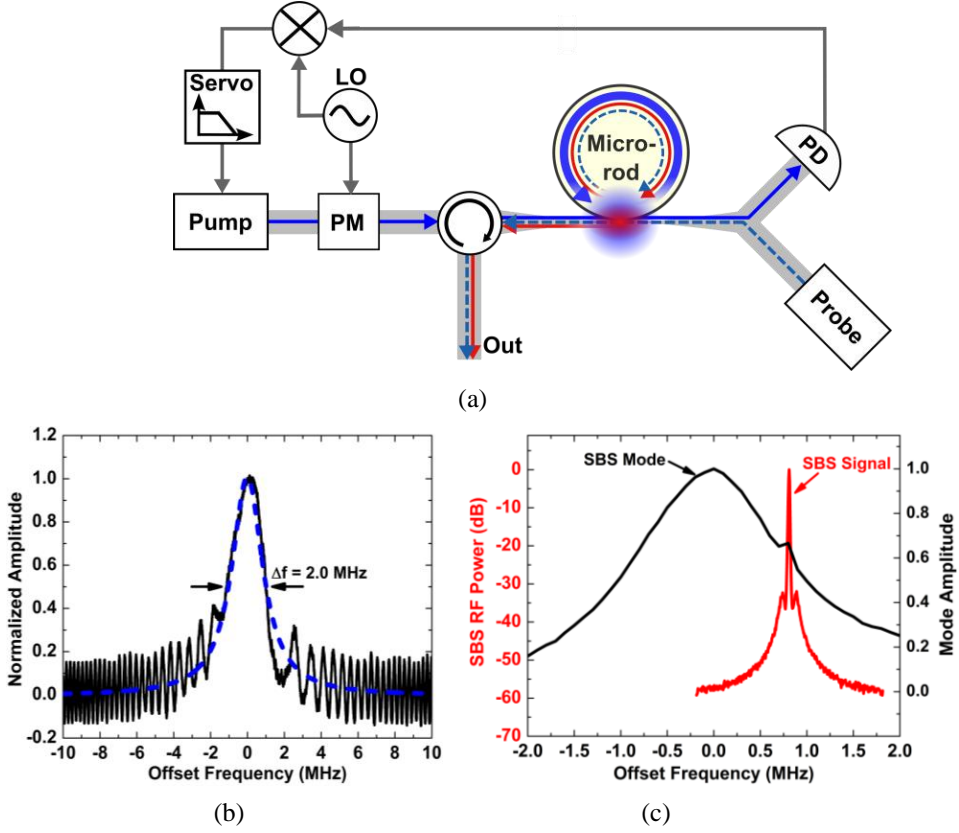

**Figure 2.** Measurement of the SBS laser operating parameters. (a) Configuration of the measurement setup used to determine the loaded Q of the microrod. (b) Profile of the pump mode illustrating zero cavity detuning and also a 2.0 MHz Lorentzian width. (c) Profile of the SBS mode illustrating a SBS detuning of 810 kHz.

In Eq. (1), $\sigma_F$ and $\sigma_B$ represent the detuning of the pump and SBS waves from their respective cavity modes, $\Gamma_b$ is the loss rate of the phonon mode, $\tau_B$ is the lifetime of the SBS mode, and $\left(\Omega_b^2 - \Omega^2\right)/2\Omega$ represents the detuning of the density wave $(\Omega)$ from the acoustic mode $(\Omega_b)$ that arises from energy and momentum conservation constraints. $\sigma_\rho$ represents the additional detuning of the density wave due to the need to balance the detuning of the pump and SBS waves in the steady state. As can be verified from Eq. (1), energy conservation enforces that the mode detunings satisfy $\sigma_B + \sigma_\rho = \sigma_F$. Formally, Eq. (1) states that the detunings of the SBS and density waves arise due to a combination of the pump detuning ($\sigma_F$) and also the phase shifts induced by operating off the SBS gain resonance $\left[\left(\Omega_b^2 - \Omega^2\right)/2\Omega\right]$. However, note that the detunings of $\sigma_\rho$ and $\sigma_B$ are asymmetric depending on the response rates of the SBS ($\tau_B$) and phonon ($\Gamma_b$) modes.

From the measurements of Fig. 2, we find that $\sigma_F = 0$ and $\sigma_B = 2\pi \times 8.1 \times 10^5$ rad/s. Furthermore, since the transfer of pump frequency noise to the SBS frequency noise depends on the response rates of the SBS and phonon modes to the pump perturbation governed by $1/\left(1+\Gamma_b \tau_B\right)^2$ [12, 30], our measurements of this noise transfer determine the operation state of the SBS laser in Eq. (1). In our experiment, we find the SBS fluctuations to be reduced by a factor of 330 compared to the pump fluctuations, and thus we determine that $1+\Gamma_b \tau_B = 18.2$. This measurement was performed by phase modulating the pump with a constant tone at higher frequencies beyond the system's thermal response (50 kHz) and measuring the resulting frequency noise on both the pump and SBS signals. From the first equation of Eq. (1), this yields a detuning of the phonon mode with respect to the SBS gain maximum of 14.7 MHz towards the lower frequency side. Finally, from the second equation of Eq. (1), we find the density wave is detuned by 810 kHz to the red of the SBS gain maximum and is detuned by 13.9 MHz to the blue side of the phonon mode. Beyond the detuning of the phonon wave, we can also calculate the linewidth of the phonon mode using both the noise transfer ratio and the measurement of the SBS mode linewidth. From Figs. 2b and c, we determine the SBS mode linewidth to be ~2 MHz, and thus calculate that $\tau_B = 80$ ns. Using this information, we find the phonon mode linewidth to be $\Gamma_b = 2\pi \times 34 \times 10^6$ rad/s or 34 MHz, which agrees with other measurements of the phonon mode linewidth found in Ref. [21]. Lastly, because of the phase coherence intrinsic to the SBS nonlinearity, the phases of the pump $(\phi_F)$, SBS $(\phi_B)$, and phonon $(\phi_\rho)$ waves all share a direct phase relation with one another. When the SBS gain is maximum, $\phi_F - \phi_B - \phi_\rho = \pi/2$, and when the SBS gain is all consumed to generate phase rotation (zero gain), $\phi_F - \phi_B - \phi_\rho = 0$ or $\pi$. Using the relation $\tan(\phi_F - \phi_B - \phi_\rho) = -1/2\sigma_B \tau_B$ [22], we find $\phi_F - \phi_B - \phi_\rho = 2.25$ thus indicating that a fraction of the SBS nonlinearity is used to compensate for phase rotations or mode detuning in our system. This detailed analysis of the SBS laser equips us with the foundation necessary for future studies of the relation between SBS laser noise and microresonator detuning.

*C. SBS Microcavity Thermal Response*

The absorption of light coupled into a microresonator results in the generation of heat, which induces a shift in the cavity resonances through both the thermal expansion of the cavity and also through the temperature dependence of the refractive index. This dependence on temperature connects the pump intensity to the frequency of the optical mode and, as we will see later, represents the major source of amplitude to frequency noise conversion in the SBS laser. Thermorefractive noise [31, 32] is another noise source that becomes dominant at low frequencies, in which fundamental temperature fluctuations induce fluctuations in the resonance frequencies. We are interested in minimizing both of these pathways for noise through the use of a fused-silica microrod resonator, which exhibits larger optical mode volume compared to conventional integrated silica [13] or silicon nitride [3] microresonators. The larger mode volume increases the thermal time constant of the system to the point where the cavity can no longer respond to faster noise fluctuations. This process then causes the thermally-driven noise to average down and to decrease in value.

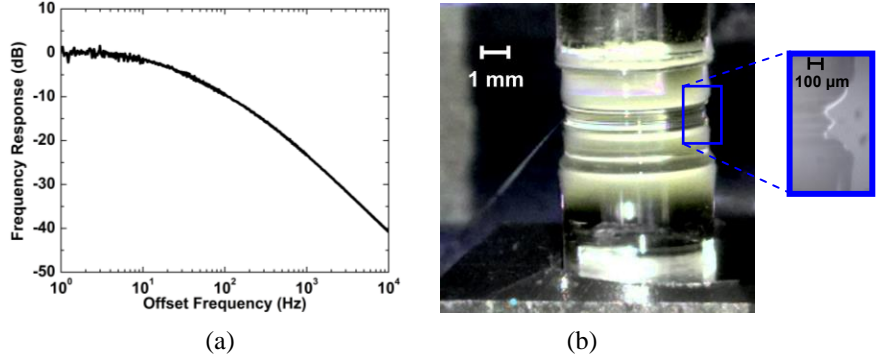

(a)          (b)

**Figure 3.** Thermal properties of the SBS microresonator. (a) Thermal response of the SBS microrod indicating a 3 dB bandwidth of 18 Hz. (b) Photograph of the SBS microrod resonator having a fabricated resonator minor diameter of 170 μm.

Figure 3a shows the measured thermal response of the microrod resonator. The measurement was performed by first stabilizing the pump laser to the cavity resonance and then subsequently modulating the amplitude of the signal coupled into the resonator via the SOA of Fig. 1a. By comparing the resulting frequency fluctuations (referenced against an independent 1-Hz linewidth cavity-stabilized laser) to the amplitude modulated input, we determine the underlying thermal response governing the coupling of pump intensity to mode frequency. The measurement of Fig. 3a indicates a thermal time constant of 18 Hz and is nearly two orders of magnitude slower than the thermal time constant of the silica microdisk resonators used in Ref. [14]. Beyond 10 kHz offset frequency, the measured thermal response becomes limited by the noise of our reference cavity-stabilized laser. The corresponding photograph of the microrod is shown in Fig. 3b. We estimate the microrod's mode-field diameter to be in the range of 100 μm.

*D. SBS Laser Noise*

The long thermal time constant of the microrod resonator combined with its low optical loss enable the microrod to support the generation of nonlinear optics with exceptional noise properties and low requirements for pump power. However, since the SBS generation process is dependent on the intracavity intensity, there exists a limit as to how large the mode volume can be made before the laser can no longer reach threshold. For our SBS laser, the threshold for lasing is reached with 6.3 mW of pump power at the tapered fiber input, and thus we are able to still achieve low operating powers due to the large slope efficiency (69 %) of the laser. For example, with 11.7 mW of pump, the SBS output is 3.7 mW.

In order to assess the performance of our microrod SBS laser, we measure its noise in terms of fluctuations in both intensity and frequency. Fig. 4a shows the relative intensity noise (RIN) of the pump and SBS lasers. We find the SBS RIN (red) to be above that of the pump laser at the microresonator input (blue). For frequencies larger than 1 kHz, the SBS RIN is white and dominated by the intrinsic noise generated by the SBS gain process. At lower frequencies, we believe the SBS RIN to be dominated by technical or excess noise, while at higher frequencies beyond 1.3 MHz, the RIN rolls off with the cavity response. We note that although the measured SBS RIN is above that of the pump laser, this RIN can be readily reduced to levels below the pump RIN using the combination of a saturated optical amplifier and/or an intensity servo [14]. Fig. 4a also shows the SBS RIN when intensity noise is intentionally injected onto the pump through modulation of the SOA power at the microresonator input. Since the intracavity pump power becomes clamped at threshold, the modulated power is transferred over to the SBS power as intensity noise. This noise has a bandwidth of 100 kHz due to limitations in the SOA modulation response and will be important for our discussion later when we analyze the fundamental performance limitations of the SBS laser.

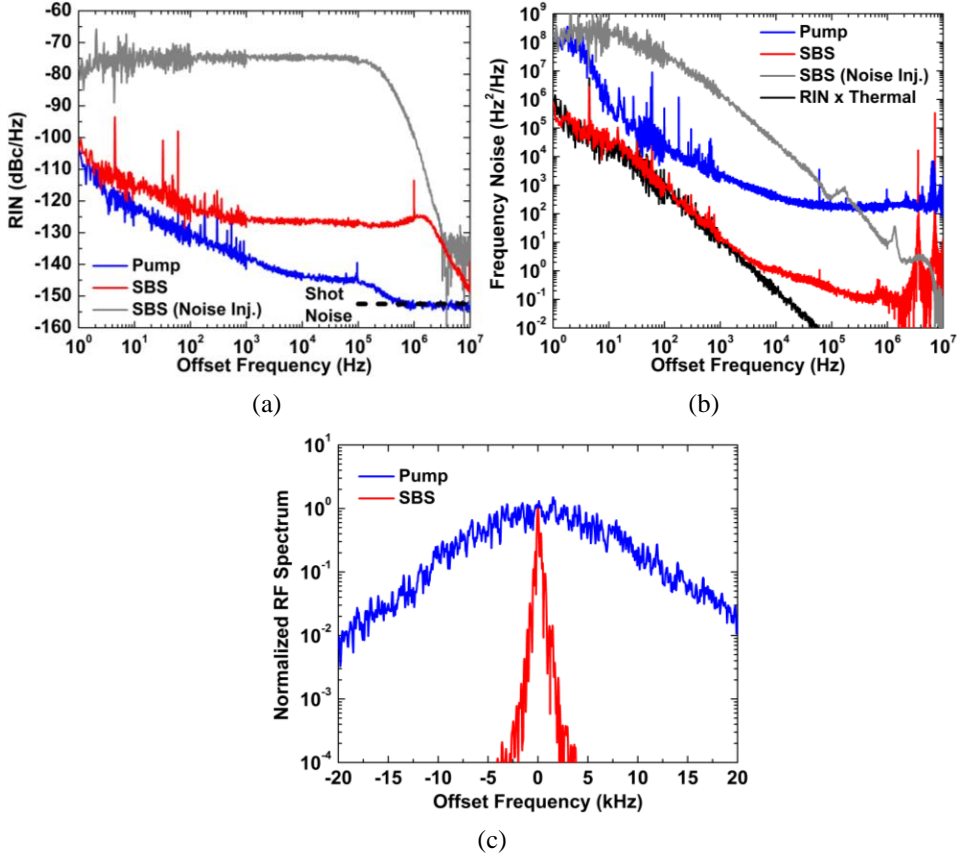

**Figure 4.** SBS Microrod laser noise characteristics. (a) RIN of the pump laser (blue), the SBS laser (red), and the SBS laser injected with broadband white noise (dark gray). (b) Frequency noise corresponding to the same laser configurations. The noise conversion from SBS laser RIN to SBS frequency noise is also provided (black). (c) RF spectrum of the pump and SBS lasers measured by a cavity-stabilized laser showing a SBS laser linewidth of 240 Hz. The resolution and video bandwidth were 100 Hz and 50 Hz, respectively.

We next measure the frequency noise performance of the SBS laser (Fig. 4b) along with the noise of the integrated planar external-cavity diode laser that we use as the pump. We perform this measurement using a combination of techniques that captures the noise across both low and high offset frequencies. At low offset frequencies, we directly heterodyne our SBS microrod laser with a 1 Hz linewidth cavity-stabilized fiber laser and detect the resulting frequency fluctuations using a frequency-to-voltage converter. However, at higher offset frequencies where the cavity-stabilized laser exhibits larger noise due to limitations in the locking bandwidth, we instead use a Mach-Zehnder interferometer [33] as a frequency discriminator with delay lengths ranging from 60 m to 550 m. We find the pump frequency noise to be 10 dB – 30 dB above the SBS noise across the entire range of the measured spectrum. At 10 Hz, the SBS frequency noise reaches $2\times10^4$ Hz$^2$/Hz and is nearly two orders of magnitude below that of the pump laser and also that of the SBS microdisk laser [14]. This improvement is a direct result of the microrod's longer thermal time constant (Fig. 3a) in comparison to that of the microdisk, which results in an averaging over the thermal fluctuations of the system. The SBS frequency noise reaches a floor of 0.1 $Hz^2$/Hz beyond 300 kHz offset, which is below the Mach-Zehnder measurement sensitivity for delay lengths below 60 m. Note that the spurs near 3.5 MHz and 7 MHz are due to the transfer function of the Mach-Zehnder measurement for ~60 m of fiber delay.

To assess the limitations in the SBS laser's performance, we again intentionally inject the same amount of intensity noise as in the case of Fig. 4a via the SOA and monitor the response of the frequency noise. Note that since the noise bandwidth is 100 kHz, the corresponding noise beyond 100 kHz in Fig. 4b is due to a combination of the servo peak of the 1 Hz linewidth cavity-stabilized laser at 150 kHz, the relaxation oscillation of the laser at 1.4 MHz, and the rolloff of the frequency-to-voltage converter beyond 5 MHz. The ratio of the measured noise spectra characterizes the conversion of amplitude noise to frequency noise in the SBS laser and follows the shape of the cavity thermal response (Fig. 3a), as one would expect. We multiply this ratio across frequency with the measured RIN of the SBS laser and find the resulting noise to be nearly identical to the frequency noise of the SBS laser. The similarity between the two provides strong evidence that the current limitations of the SBS laser stem from fluctuations in amplitude.

The RF spectrum of the beat note between the SBS laser and the cavity-stabilized laser (Fig. 4c) provides an additional independent measure of the overall noise performance of the SBS laser. For this measurement, we used a resolution bandwidth of 100 Hz with a corresponding analyzer sweep time of 0.028 s. The spectrum of the pump laser is provided in Fig. 4c for reference. Since the cavity-stabilized laser's linewidth is much narrower than that of the SBS laser, the resulting spectrum is a true measure of the SBS laser's lineshape. By integrating the power within the SBS laser lineshape, we determine its half-power linewidth to be 240 Hz. This agrees closely with the linewidth found from integrating the frequency noise of the SBS laser in Fig. 4b [34], which we also determine to be 240 Hz. The corresponding half-power linewidth of the pump is 7.9 kHz and is thus much larger than that of the SBS laser, as one would expect from the frequency noise spectra of Fig. 4b.

In Figs. 4a and b, the coupling between amplitude and frequency through the microresonator's thermal response provides insight regarding the fundamental noise limitations of the SBS laser. Here, a fluctuation in the intracavity power produces a change in the resonator's intensity absorption and thus ultimately results in a fluctuation of the system's internal temperature. This change in temperature shifts the modes of the resonator, thus provoking a frequency shift in response to the fluctuations of amplitude. The conversion of amplitude noise to frequency noise is captured by the microresonator's thermal response in Fig. 3a and also by the response of the injected noise in Figs. 4a and b. To show the correlation between the two, we provide the measured frequency response between the SBS laser RIN and frequency noise in Fig. 5a alongside the cavity thermal response. The magnitude of this response falls off near 10 Hz offset frequency, in exact agreement with the thermal time constant of the microrod. Furthermore, the phase of the response decreases steadily with offset frequency and is non-random, thus indicating that the SBS intensity and frequency are correlated over the spectrum.

The correspondence between SBS intensity and frequency fluctuations enables us to correct for frequency jitter through a servo of the intensity noise. We show here the results of our noise correction scheme which utilizes the detection of the SBS laser RIN as a feedback signal to control the SOA output power before the microresonator. Since the intracavity pump power is nominally clamped at the SBS lasing threshold [22], the changes in the pump power all go towards cancelling the fluctuations in the SBS laser intensity. In addition to reducing the SBS frequency noise, the servo has the added benefit of simultaneously suppressing the SBS laser intensity fluctuations. The resulting SBS RIN after servoing the laser intensity noise is shown in Fig. 5b. In comparison to the uncorrected RIN, the servo improves the SBS laser's intensity fluctuations by 10 dB – 20 dB up to offset frequencies of a few kilohertz.

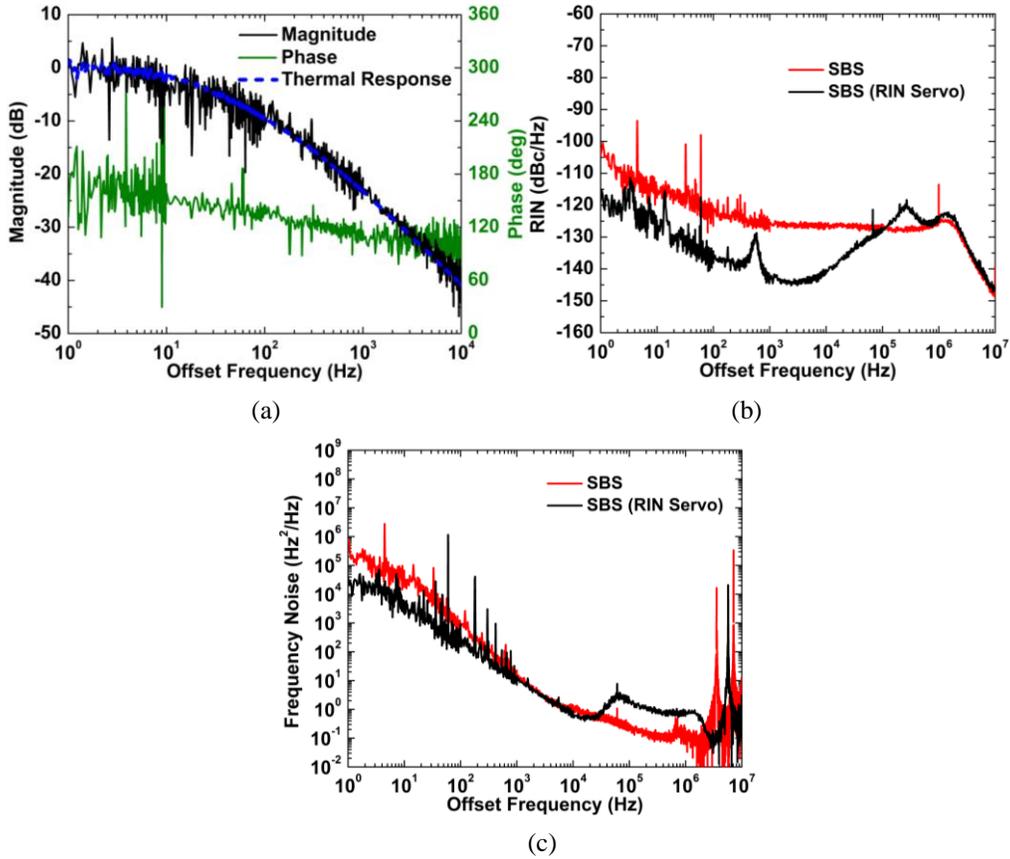

**Figure 5.** SBS microrod laser noise feedback. (a) Frequency response between SBS RIN and SBS frequency noise indicating a rolloff that matches the microrod thermal response. (b) RIN of the SBS laser showing 10 dB – 20 dB noise improvement upon servo control of the SBS intensity fluctuations. (c) SBS laser frequency noise showing an improvement at low offset frequencies as a result of the RIN servo.

Fig. 5c shows the improvement in the SBS laser's frequency noise due to feedback provided by the RIN servo. At lower offset frequencies near ~1 Hz, the frequency noise improves by an order of magnitude, while at higher frequencies near 100 kHz, the frequency noise increases slightly. From our tests, this increase in noise appears to be due to the interplay between slight changes of our cavity mode/SBS laser state and the microresonator Pound-Drever-Hall servo operation rather than to the addition of the RIN servo. This can be readily verified by removing the RIN servo and observing the resulting SBS laser noise level. Finally, the residual spur at 5.7 MHz is again due to an artifact of our Mach-Zehnder measurement performed with a delay length of 35 m. By integrating the spectrum of the frequency noise, we determine that the SBS laser linewidth improves slightly with servo feedback down to a linewidth of 225 Hz. As can be observed from Fig. 5, most of the improvement in noise occurs for offset frequencies below 100 Hz and thus contributes negligibly to the spectral width at the 3 dB point.

## 4. Summary

We showed that the thermal fluctuation of a microresonator, which is an important source of noise at low frequencies for all microcavity devices, becomes averaged and reduced by the thermal response of the system. For this reason, we explore the use of a microrod resonator with a thermal bandwidth of 18 Hz as a medium for ultralow noise lasing via the stimulated Brillouin scattering optical nonlinearity. The SBS microrod laser demonstrates two orders of magnitude lower noise compared to similar SBS lasers that use a chip-integrated resonator platform and exhibits a measured half-power linewidth of 240 Hz. At this level, we determine from measurements of the noise correlation that our SBS laser noise is still limited by the conversion of intensity noise to frequency noise through the thermal response of the microrod. This coupling between amplitude and frequency sheds light on the intrinsic noise limits of microresonator devices and provides a pathway to use simple measurements of intensity as the means to correct for fluctuations in frequency. The techniques that we demonstrated have the advantage of avoiding the necessity for ultrastable frequency references that would otherwise complicate the system and/or increase its size, weight, and power consumption.

## Acknowledgments

We thank Dr. Marco Schioppo, Dr. Gabe Ycas, and Prof. Kerry J. Vahala for their valuable comments on this manuscript. This work was funded by NIST and the DARPA PULSE Program. WL acknowledges support from the NRC/NAS. DC acknowledges support from the NSF GRFP under Grant No. DGE 1144083. This work is a contribution of the US Government and is not subject to copyright in the US.